\documentclass{article}

\usepackage[final]{neurips_2019}

\usepackage{times}
\usepackage{latexsym}
\usepackage{graphicx}
\usepackage{amsmath}
\usepackage{pgfplots}
\usepackage{natbib}
\pgfplotsset{width=0.7\textwidth}

\usepackage{url}
\usepackage{comment}

\title{Automatic Story Construction from News Articles in an Online Fashion}

\author{%
  \"{O}zg\"{u}r Can\\
  Dept. of Computer Engineering\\
  \.{I}zmir Institute of Technology\\
  \.{I}zmir, Turkey \\
  \texttt{ozgurcan.can@gmail.com} \\
  \And
  Selma Tekir \\
  Dept. of Computer Engineering\\
  \.{I}zmir Institute of Technology\\
  \.{I}zmir, Turkey \\
  \texttt{selmatekir@iyte.edu.tr} \\
}

\date{}
\pgfplotsset{compat=1.14}
\begin{document}

\maketitle

\begin{abstract}
This paper presents a novel story construction system to track the evolution of stories in an online fashion. The proposed system uses a novel sliding window solution, named Inching Window, allowing the processing of each new data element on-the-fly. To assign a new data element into a community in a fast and memory-efficient manner, we apply the modularity maximization idea of Louvain method on-the-fly. As part of the experimental validation, we provide step by step construction of a meaningful news story and support the case with a set of visualizations.
\end{abstract}


\section{Introduction}
Every day, thousands of local and global news become online. Each arriving news piece gets its meaning through its connections with some other news. Understanding the present situation requires an analysis in light of past. Thus, organizing news content in a coherent set of articles to form a story is a fundamental requirement.

Automatic story construction is a challenging task subject to some key issues. Firstly,  data may come from multiple sources and could include overlaps. Moreover, it must be processed in an online fashion. Time-span may become another difficulty as long ranging stories must be supported as well as short-ranging ones. Visualization is another concern because stories gain meaning in the eyes of the beholder.

\begin{figure}
  \centering
  \includegraphics[width=1\textwidth]{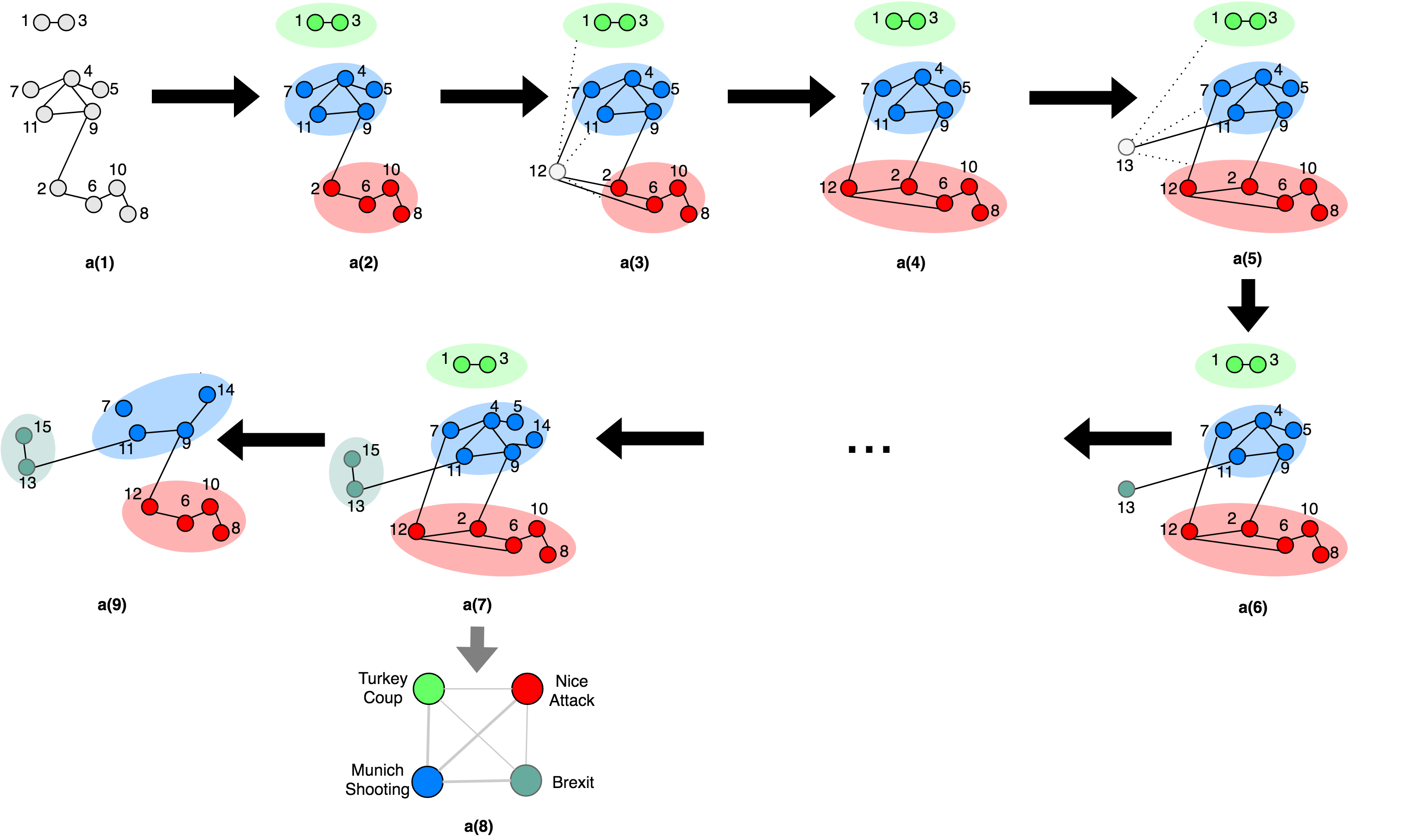}
  \caption{A sample evolving set of stories using our proposed system.}
  \label{fig:sample_corpus_execution}
\end{figure}

Topic Detection and Tracking (TDT) \cite{Allan2002TopicDA} covers identifying and following events from a constantly arriving stream of text from multiple sources. TDT defines an event as something that happens at a particular time and place and aims to group related documents that discuss the same event. At a higher conceptual level, a topic is defined as a theme that is related with a set of events. Topics are combined to form stories that represent a coherent structure of news. To illustrate story structures from a document stream, a sample mini-corpus of $15$ news articles on four stories; \textbf{Turkey's Coup}, \textbf{Nice Attack}, \textbf{Munich Shooting} and \textbf{Brexit} is processed using our proposed system as depicted in Figure \ref{fig:sample_corpus_execution}. As can be seen from the figure, \textbf{Turkey Coup} story appears at the beginning of the period. At the final stages, it disappears. While \textbf{Nice Attack} and \textbf{Munich Shooting} remain steady all along the time period, \textbf{Brexit} becomes evident eventually.

In literature, story structures are represented in different ways, as story graphs \cite{yang2009}, \cite{subasic2013}, information maps \cite{shahaf2012}, \cite{shahaf2015} or story trees \cite{liu2017}.

In their study, \citet{laban2017} propose a sliding window-based approach to story construction. Articles are represented with a set of TF-IDF filtered keywords. Their system operates based on three phenomena: Linking, splitting, and merging.

\citet{ansah2019} present a novel graph-based timeline summarization system, named StoryGraph, to track the evolution of the stories from online Twitter communities.

\citet{hu2017} propose a word embeddings-based document representation method and an online event detection algorithm, which uses time slicing and event merging. 

In this work, we propose a system to generate continuously updated news stories. The main contributions of this work can be summarized as follows:

\begin{enumerate}
\itemsep0em
    \item Novel way to represent stories as vectors.
    \item Novel sliding window approach (Inching Window) that reduces the time complexity of processing each item in the context of the recent data.
    \item Online clustering method (Louvain on-the-fly), which can automatically detect new stories from a continuous stream.
\end{enumerate}

\section{Updating News Stories}

\subsection{Document Representation}\label{sec:document-representation}
We use doc2vec \cite{le2014} embeddings to represent documents.

\subsection{The Clustering Algorithm}\label{sec:clustering-algorithm}


In this work, we represent each news article as a node in a fully connected undirected weighted network, where the edge weights between the nodes are set based on the cosine similarities between the document vectors.

 To detect the stories, we use the Louvain method \cite{blondel2008}. The Louvain community detection method is an algorithm for finding communities in the network by trying to maximize the modularity in a repeated process. One of the major advantages of the Louvain method is that it reveals a hierarchy of the communities at different scales and this hierarchical perspective helps us understand the story construction and observe root stories.
 
\subsection{Story Representation}\label{sec:story-representation}

Learned document representations can capture syntactic and semantic regularities and they also preserve these regularities on algebraic operations. Thus, in this work, we propose to use this feature to come up with story representations. To be specific, we add up all the document vectors in the story to calculate the story vector.

\subsection{Online Clustering}\label{sec:online-clustering}

Stream clustering algorithms are used for extracting useful knowledge from data streams. 

One common technique for online clustering is evaluating the new data point only over sliding windows of recent data. For instance, only data from the last week can be used to cluster new data instead of querying over entire past history. Sliding window approach emphasizes only the recent data, and for many applications, recent data are more important and relevant than the old one. 

Majority of the news articles belonging to the same story are published in a dense way in a short period of time. Considering this fact, many existing works on online news story detection feed the news stream in the chronological order to their systems and use sliding window method, and time-sensitive queries to cluster the documents \cite{silva2013}. In such systems, defining an optimal sliding window interval is a challenge. 

\subsection{Inching Window}\label{sec:inching-window}
\begin{figure}
  \centering
  \includegraphics[width=0.7\linewidth]{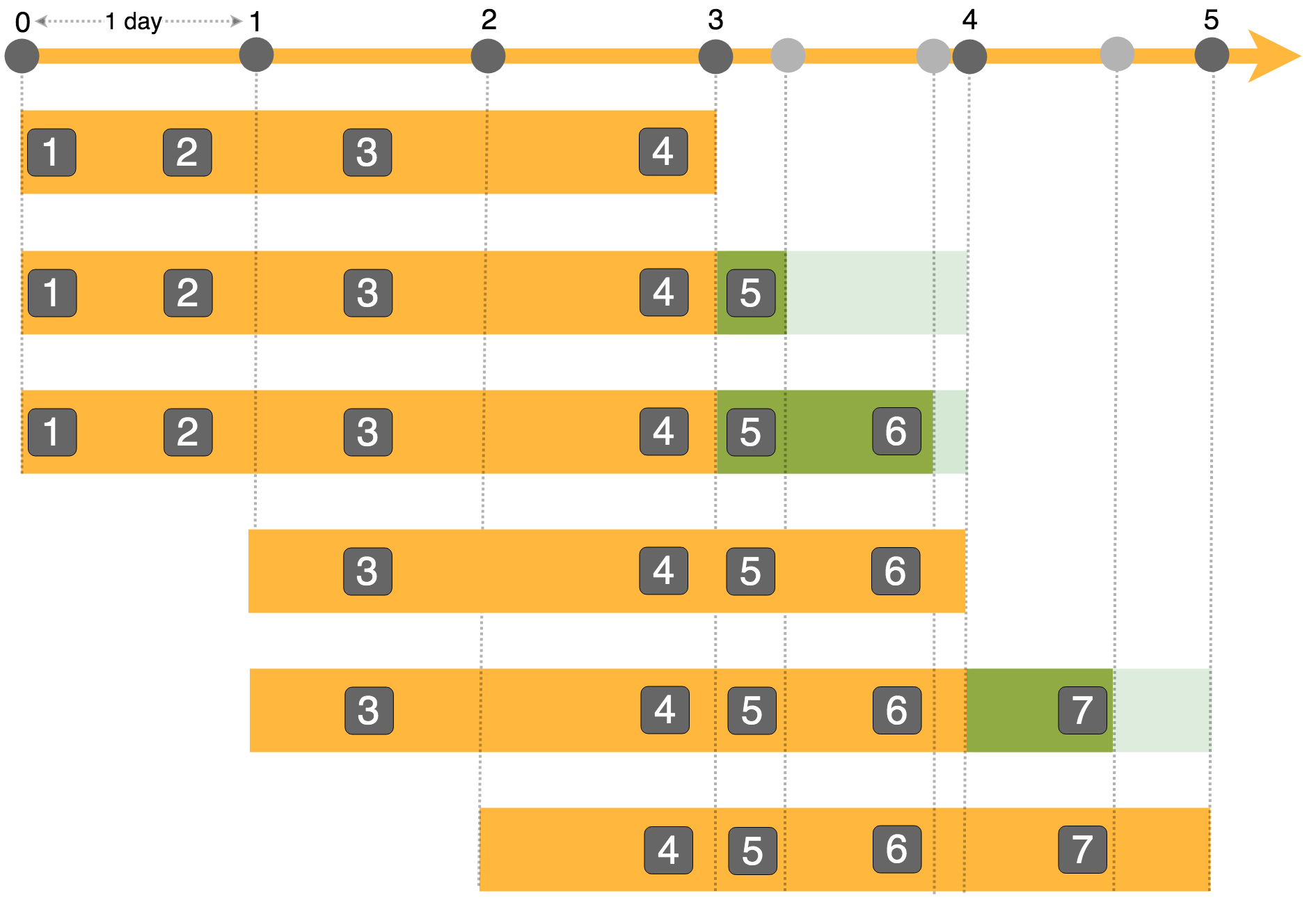}
  \caption{Illustration of the inching window process.}
  \label{fig:inching_window}
\end{figure}

Generally, the main concern of a data streaming system is processing new data elements on-the-fly, but processing with respect to a time interval requires gathering the data until interval condition is met. For instance, a sliding window of five days with a one-day sliding interval requires to wait until the end of the last day in the window to gather all data. To overcome this issue, we propose a novel sliding window technique, named \textit{"Inching Window"} as the window slides like an inchworm. Inchworms move in a characteristic \textit{inching} or \textit{looping} gait by extending the front part of the body and bringing the rear up to meet it" \footnote{https://www.merriam-webster.com/dictionary/inchworm}. An \textit{inching window} moves in a similar fashion. Given a window size of $3$ days, and a sliding interval of $1$ day, similar to sliding window, \textit{inching window} groups the incoming data in batches every $3$ day, but rather than sliding after processing the batch, it continues to process each incoming data one by one until the sliding interval ($1$ day), then it slides. This \textit{inching window} process is illustrated in Figure~\ref{fig:inching_window}.

The main advantage of the \textit{inching window} approach over \textit{sliding window} is that \textit{inching window} enables processing of the new data elements on-the-fly within the recent data context.

After feeding a news stream in the chronological order to an \textit{inching window} set up, we propose an online story detection system that is composed of three main steps: Topic creation, on-the-fly document clustering, and story construction.

\subsection{Topic Creation}\label{sec:topic-creation}
A set of news articles in a small time interval (e.g. $5$ days) have a potential to create a densely-connected local group around a \textit{topic} which, can then be merged in a \textit{story}. To detect the topics in a window, we initiate a news article network and apply the clustering algorithm explained in Section \ref{sec:clustering-algorithm}. The resultant communities are topics and they can be represented with topic vectors, which are computed by adding all document vectors belonging to the topic.

\subsection{On-The-Fly Document Clustering}\label{sec:clustering-on-the-fly}
After the initial inching window, news articles are received one-by-one until the specified window interval. This enables us to process a newcomer article without waiting for the end of the day. However, it is not computationally feasible to apply the clustering algorithm again and again for each newcomer article. Furthermore, it is often acceptable to assign a newcomer article to the most suitable community in the given context, and later on, review the temporary assignments and then make permanent assignments.

\begin{figure}
    \centering
    \includegraphics[width=.6\linewidth]{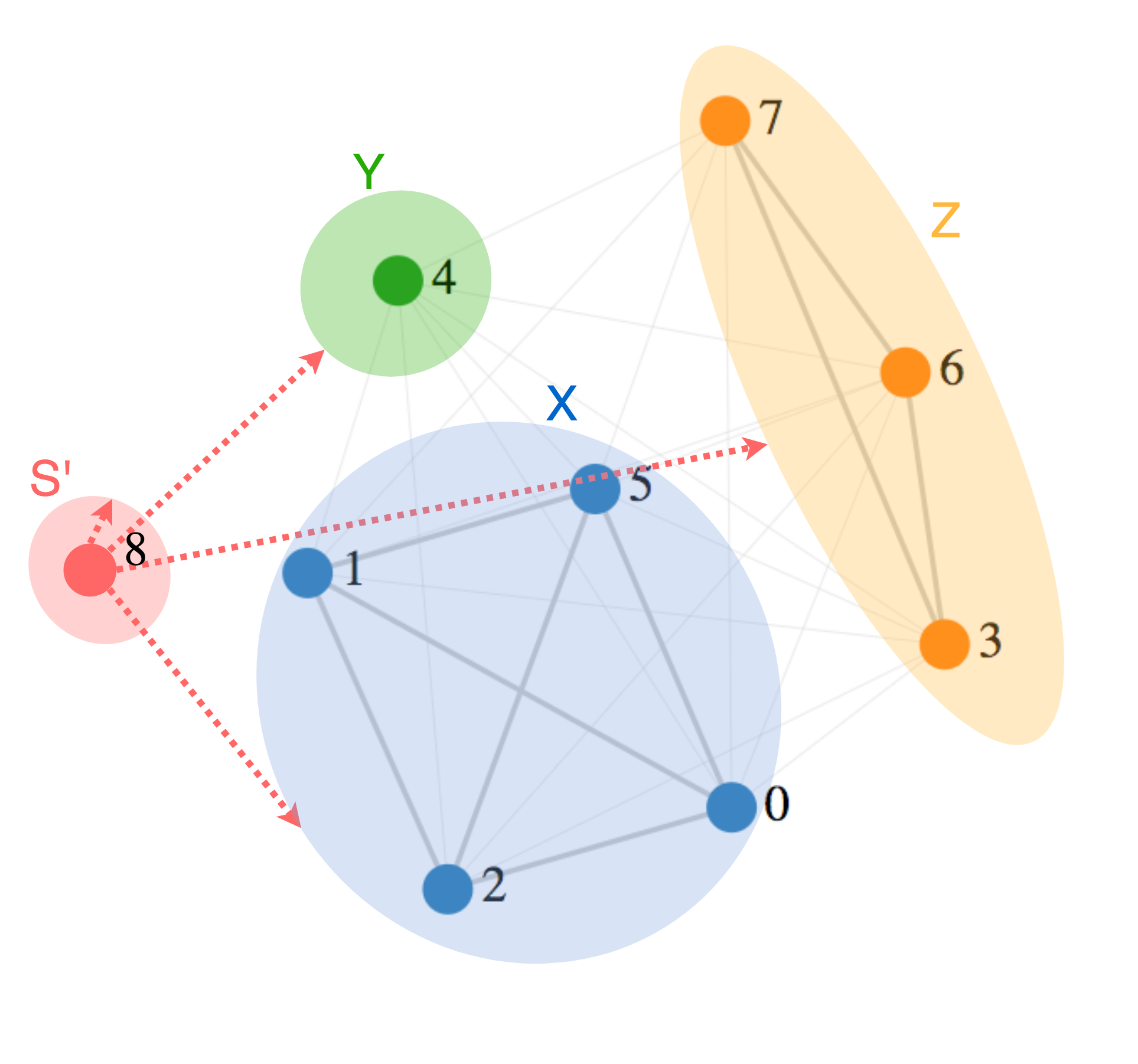}
    \caption{On-the-fly clustering process.}
    \label{fig:figure_on_the_fly_clustering}
\end{figure}

The clustering algorithm that we use in Section \ref{sec:clustering-algorithm}, relies on Louvain community detection method \cite{blondel2008} that tries to identify communities in a network by repeating optimization steps iteratively until a maximum value of modularity is attained. 
To assign a new document into a community in a fast and memory-efficient manner, we propose to compare modularity changes in the network after assigning the new document into one of the existing communities or creating a new community for the document. The on-the-fly news clustering process is illustrated in Figure \ref{fig:figure_on_the_fly_clustering}. Given a network with communities found ($X$-blue, $Y$-green, and $Z$-yellow), for the newcomer $Article_8$ a new community named $Community_S'$ is created. Then, we measure modularities on the graph for the scenarios where $Article_8$ joins $Community_X$ or $Community_Y$ or $Community_Z$. Then, we select the scenario where the graph modularity is maximized. For the given sample, we found that community X maximizes the modularity, so $Article_8$ is assigned to $Community_X$. 

Before sliding a window, we apply the clustering algorithm all over the window and override the temporary assignments. 

\subsection{Story Construction}\label{sec:story-construction}
A \textit{topic} that we detect might be; 1. an initial seed of a new story, 2. belonging to an active story, 3. a continuation of a story that is interrupted in time, for instance, a crucial piece of evidence for \textit{Malaysia Airlines Flight 370 Disappearance} is found after years.

To properly address these characteristics of a topic, we propose to create another fully connected network where the nodes are stories detected, and the edges' weights are cosine similarities between the nodes. First, we represent topics found in a window with the sum of the vectors representation of the articles belonging to that topic. Then, we added the topics as a node to the story network that we initiated. Then, we apply the community detection algorithm. Running community detection over the story network creates interesting dynamics: 

\begin{enumerate}
\itemsep0em
    \item A topic might create a community with an existing story in the network. In this case, we merge the topic with the story, then remove the topic node from the network. Merging is executed by summing up the story vector and the topic vector after handling the document migrations from one story to another. Sliding windows overlap in time and data. Thus, a topic found in a window can have the same document with a story found in the previous window. Thus, when a story and topic is merged, if a document is already belonging to another story, it should be removed from that story first.
    \item A topic might create a new individual community. After handling the document migrations, the topic is directly casted into a story.
    \item A topic might create a community with another topic found in the same window. Communities are decided based on the modularity of the network. Two document communities found in the local article network context might be merged into one in the global story network. Merging is executed in the same fashion with topic-story merging.
    \item  An existing story might create a community with another story. This case rarely happens, but as the stories grow in time, their relations with each other evolve as well. Thus, two distinct stories might start to have connections and finally be merged. Two stories are merged on the oldest one by directly summing up their vectors. 
\end{enumerate}

In our system, stories do not need to keep individual document data. In creating a story from a topic, keeping only the document ids and sum of the document vectors to represent the story is sufficient. Document data and their individual vector representations can be deleted from the memory. \\

\section{Experimental Results}

To illustrate the story construction process, we executed our proposed method in a sample mini-corpus of 15 news articles between 24.06.2016 and 29.06.2016. 
The corpus consists of news articles on four stories; \textit{Turkey's Coup}, \textit{Nice Attack}, \textit{Munich Shooting} and \textit{Brexit}. We set up an inching window for 4 days with the inching interval of 1 day (Figure \ref{fig:sample_corpus_execution}). In Figure \ref{fig:sample_corpus_execution}, in order to simplify the tracking of the stories, we do not depict the network as a fully connected one. 

\begin{enumerate}
\itemsep0em
    \item Subfigure a(1) shows the network of the first window articles between 24.06.2016 and 28.06.2016.
    \item In the Subfigure a(2), the clustering algorithm is run to detect the topics.
    \item Forward moving process of the inching window is started in Subfigure a(3). To assign the newcomer \textit{$Article\textsubscript{12}$} to a community, the graph modularity is iteratively checked for each possible community assignments.
    \item In the Subfigure a(4),  \textit{$Article\textsubscript{12}$} is assigned to the community that produces the maximum modularity.
    \item Subfigure a(5) and Subfigure a(6) apply the same procedure as the Subfigure a(3) and Subfigure a(4) for the  \textit{$Article\textsubscript{13}$}, but this time, creating a new community maximizes the graph modularity.
    \item On-the-fly clustering is executed for each new document until \textit{$Article\textsubscript{15}$}, which is the latest article of the window. This last step of the window is shown in Subfigure a(7).
    \item Topic clustering algorithm is run again to review and correct the on-the-fly community assignment in Subfigure a(8).
    \item In the Subfigure a(8), the final set of topics is added to the story network and proposed flow for the story construction is executed.
	\item In the Subfigure a(9), we shrink the inching window by removing the first day of the window; 24.06.2016.
\end{enumerate}

In order to further evaluate the performance of the proposed system, we created a sample corpus of $400$ documents which contains a short-ranging story, a long-ranging story, and a story that is interrupted in time and fed the set of documents as a stream. As a result, we achieved to get high $F1 (0.829)$ and $NMI (0.823)$ scores  that indicate a good clustering performance.

\section{Conclusion}

In this work, we present a method for aggregating news articles into topics and then merge them into story clusters in an online fashion.

To the best of our knowledge, this is the first work that uses doc2vec \cite{le2014} for news article representation in topic detection and story construction. To perform the task in an online fashion, we present a modified sliding window approach called \textit{"Inching Window"}. Moreover, in accordance with the Inching Window concept, we use the modularity maximization idea of Louvain method to perform on-the-fly clustering. We will publish our code so that the research community can build on top of our work.

\section{Future Work}\label{sec:future-work}
Revealing communities in document article network allows for analysis of relations between individual articles and classifications of the objects based on their characteristics. In some real-life networks such as social networks, objects can simultaneously belong to multiple communities at once. These communities are called overlapping communities. In this study, we follow the general traditional sense that a news article can belong to only one news story at a time. However, analysis of the overlapping news stories might create interesting dynamics. We plan to change switch Louvain community detection algorithm with an overlapping community detection algorithm and evaluate the performance of the system.

Although the focus of the work is to find stories from the news articles, we plan to test the proposed system with different types of data, such as social media news. 

In this work, we evaluate the accuracy with a manually collected document set. Real world data might have more noise and different characteristics. We plan to evaluate the accuracy of the story creation algorithm in a large annotated corpus.

\bibliography{arxiv}

\end{document}